# BRATsynthetic: Text De-identification using a Markov Chain Replacement Strategy for Surrogate Personal Identifying Information

John D. Osborne, PhD[1,] Tobias O'Leary, B.S.[1,] Akhil Nadimpalli, Salma M. Aly. Richard E. Kennedy, M.D.

[1]University of Alabama at Birmingham, Birmingham, Alabama, United States

Corresponding author & address: John D. Osborne, 136 Tinley Harrison Tower, 1720 2nd Avenue South, Birmingham, Alabama, 35294

Contact Information:
Email: ozborn@uab.edu
Phone Number: 205.996.0736

## ABSTRACT

**Objective:** Implement and assess personal health identifying information (PHI) substitution strategies and quantify their privacy preserving benefits.

**Materials and Methods:** We implement and assess 3 different "Hiding in Plain Sight" (HIPS) strategies for PHI replacement including a standard Consistent replacement strategy, a Random replacement strategy and a novel Markov model-based strategy. We evaluate the privacy preserving benefits of these strategies on a synthetic PHI distribution and real clinical corpora from 2 different institutions using a range of false negative error rates (FNER).

**Results:** Using FNER ranging from 0.1% to 5% PHI leakage at the document level could be reduced from 27.1% to 0.1% (0.1% FNER) and from 94.2% to 57.7% (5% FNER) utilizing the Markov chain strategy versus the Consistent strategy on a corpus containing a diverse set of notes from the University of Alabama at Birmingham (UAB). The Markov chain substitution strategy also consistently outperformed the Consistent and Random substitution strategies in a MIMIC corpus of discharge summaries and on a range of synthetic clinical PHI distributions.

**Discussion:** We demonstrate that a Markov chain surrogate generation strategy substantially reduces the chance of inadvertent PHI release across a range of assumed PHI FNER and release our implementation "BRATsynthetic" on Github.

**Conclusion:** The Markov chain replacement strategy allows for the release of larger de-identified corpora at the same risk level relative to corpora released using a consistent HIPS strategy.

## BACKGROUND AND SIGNIFICANCE

Privacy considerations, legal mandates or other ethical imperatives create a scarcity of clinical text for model training and standardized evaluation for machine learning. Recent efforts to synthetically generate clinical notes have shown promise, but more research is needed before widespread use of this technology[1]. Creating synthetic or artificially generated text remains an open problem in Natural Language Processing (NLP), even with recent advances made by software such as BART[2], T5[3], GPT-3[4]. An alternative approach is to use real notes and remove or replace only personal identifying information (PII), including personal health information (PHI) in the healthcare context using a process termed "de-identification". In the United States, the governing legal framework for privacy in healthcare is the Health Insurance Portability and Accountability Act (HIPAA), which under Section 164.512 of the Privacy Rule provides both the standard and implementation guidelines for two methods of de-identification. The "safe harbor" de-identification method specifies which data elements be removed. While these elements may be targeted by de-identification software[5], this software may not replace identified PHI with surrogate text. Additionally, surrogate generation software is not freely available to clinical informatics community users[6]. Consequently, redacted text may be replaced by the category label of the PHI removed and not resemble the original text, thereby making it more difficult to train machine learning algorithms where training and test data should be similar. The challenge of creating high quality de-identified synthetic text for use with machine learning may lead even large, multi-national corporations to skip the process entirely, leading to PHI data leaks[7]. Most patient re-identification occurs when data is not de-identified according to existing standards[8].

Software has been developed to de-identify clinical text; however, cost, availability, and licensing restrictions have limited public evaluations to only a subset of software[9 10]. Software evaluated has included Amazon's Comprehend Medical [11], Clinacuity's CliniDeID[12], National Library of Medicine's Scrubber[13], NeuroNER [5] and MITRE Identification Scrubber Toolkit[14], Massachusetts Institute of Technology de-identification software[15], Emory Health Information DE-identification (HIDE) software[16] and Neuro named-entity recognition (NeuroNER). However, most of this software does not provide the option to output replacement text, which is unfortunate because PHI identification at detection time could suggest a natural realistic replacement. To our knowledge, only two of these systems provide the option for PHI surrogate generation or resynthesis. Clinacuity's CliniDeID[12] provides resynthesis in the base version at the cost of 5 cents per 5,000 characters, but this capability is not available in the free version. Resynthesis is also coupled to Clinacuity's de-identification process, so surrogate generation cannot be done on a previously de-identified data set that lacks surrogate values, such as the widely used MIMIC corpus [17]. On the other hand, MIT's de-identification software [15] (used for MIMIC corpus de-identification) has resynthesis capability, but this process is not automated. Instead, a Java user interface that allows the user to select a suitable replacement for PHI in the same format.

The lack of automated surrogate generation software is unfortunate because surrogate generation can address the "residual" PHI problem"[18] by limiting the ability of human readers to identify approximately 70% of PHI in clinical text[19]. This "hiding-in-plain-sight" (HIPS)[14] strategy relies on similarity between the original text and the synthetic, making leaked PHI inconspicuous. This means a HIPS approach is preferable from a re-identification perspective versus simply replacing PHI with the detected entity type, a common default output of PHI Named Entity Recognition (NER) software and applied to the widely distributed MIMIC corpus. However, existing HIPS protocols replace PHI consistently, such that a patient or provider name would be replaced by the same surrogate across the note. While this consistent replacement approach provides more realistic text, the presence of a single critical missed personal identifier (such as a patient name) gives an attacker the ability to identify the patient, even without complex statistical analysis. Thus, a HIPS corpus has the same single occurrence re-identification vulnerability as PHI replaced by entity names. This vulnerability could expose two-thirds of leaked PHI in an attack utilizing machine learning, thus further improvements in assuring privacy protection are warranted[19] given machine learning's increasing prevalence. Another machine learning approach that HIPS is vulnerable to are "parrot attacks", where an attacker employs a similar annotation and de-identification method on the target text as the de-identification corpus provider. The attacker then examines any false negatives to infer that PHI has been missed [20]. Thus, a HIPS approach that removed the single occurrence re-identification vulnerability would be preferable.

In this paper we implement and evaluate different HIPS surrogate replacement strategies where PHI mentions are replaced by surrogate text without the constraint that such a replacement must be consistent with previous replacements of the same type. For example, each mention of a patient name in a document could be replace by a randomly selected name in our "Random" strategy or by a name selected from a Markov chain in our "Markov" strategy. We implement these two strategies and quantify their privacy preserving benefits relative to the traditional "Consistent" surrogate replacement strategy. To our knowledge no statistical evaluation has been done on these PHI substitution strategies using either synthetic or realistic clinical data sets and we provide the first such implementation decoupled from the de-identification process.

**OBJECTIVE**

We develop BRATsynthetic to provide better privacy protections in de-identified corpora. Our goal is to create the first open-source implementation for PHI surrogate addition or "resynthesis" and describe the statistical benefits to privacy protection afforded by different surrogate substitution strategies.

**MATERIALS AND METHODS**

**Software Implementation**

BRATsynthetic uses the BRAT[21] annotation format and the associated text file (if available) to create surrogate PHI independent of the software originally used to identify that PHI. BRATsynthetic generates realistic text for 27 categories of PHI as described in the Resynthesis Elements section below. While BRATsynthetic can utilize the associated original identified text to generate better replacements, it also capable of generating surrogate text without the original file, although results are less realistic. Dates and ages are offset by a random number.

Resynthesis Elements

BRATsynthetic generates a superset of I2B2 2014 personal health information entity types**[22],** including standard HIPAA Safe Harbor Categories, as well as more specific category types under the BRATsynthetic "Replacement Type" column in Supplementary Table 1. There is a many-to-many correspondence, where some data elements map to multiple BRATsynthetic types (ex. NAMES), others are a one-to-one mapping (ex. EMAIL), and others condense different Safe Harbor Elements into a single BRATsynthetic type (ex. IDNUM). Supplementary Table 2 provides an overview of this and other label differences between Safe Harbor Elements and BRATsynthetic categories. BRATsynthetic does provide additional resynthesis capacity by the inclusion of the PROFESSION category (shared with I2B2) and the TIME category (not required either under HIPAA Safe Harbor or used by the I2B2) which we include due to the increasing availability of second and sub-second data collected from personal devices not anticipated in the original guidelines. BRATsynthetic lacks UNIQUE text resynthesis ability, where UNIQUE corresponds to a section of text that could potentially reveal personal identity such as "governor's wife" or details of unusual accidents.

Surrogate Substitution Strategies

BRATsynthetic implements 3 substitution strategies, Consistent, Random and Markov. An example substitution with 6 name entities is shown in Table 1. All 3 strategies are implemented as a simple 2-state Markov chain that can either select a new surrogate value or repeat the previous surrogate value. The initial state is always a new surrogate value, but the single state transition probability varies between strategies from 0 (Consistent, always keep the previous surrogate value), 0.5 (Markov, choose new surrogate half the time) to 1 (always select a random value). The Consistent substitution strategy always uses the same replacement value, so even a single occurrence of a false negative (FN) will be identifiable as PHI. For the Random strategy, a surrogate value is selected at random from the faker[23] library for each occurrence of each entity type and the same random value to be reused as a replacement. Should a FN occur once in the de-identified document for that entity type (leaving in the patient's real name for example) it may not be immediately obvious with the Random strategy that the real name was not just another surrogate value. This is true when the number of FN errors of that critical type is lower than the maximum number of times the same surrogate value is repeated. We refer to this latter value as the "*maximum surrogate repeat size*". The Markov substitution strategy operates in the same way as the Random strategy, but previously used surrogate values are much more likely to occur due to the 0.5 state transition probability. This generates a larger maximum number of times a surrogate value is repeated (*maximum surrogate repeat size*) for a critical entity type so it can better "mask" FN errors. A sample run of the Markov process with 2 state transition after the initial random state is shown on an entity note with 6 names in Table 1 to illustrate the state transitions and output. However, this linear replacement strategy in Table 1 would allow attackers to detect missed names in runs of synthetic names. So BRATsynthetic implements non-linear replacement strategy to disperse clusters of same surrogate names throughout the document. A patient level non-linear replacement strategy operating over multiple documents is also desirable, but not yet implemented in BRATsynthetic.

**Table 1. Substitution Strategies in BRATsynthetic**

|  | Substitution Strategy | | | |
| --- | --- | --- | --- | --- |
|  | **None** | **Consistent** | **Random** | **Markov** |
| **Original Name** | *Sandy* | *Sandy* | *Sandy* | *Sandy* |
| **Surrogate** | *ENTITY_NAME* | Sara | Kim | Sara |
| **Surrogate** | *ENTITY_NAME* | Sara | Nisha | Sara |
| **Surrogate** | *ENTITY_NAME* | Sara | Cathy | Ann |
| **Surrogate** | *ENTITY_NAME* | Sara | Maria | Maria |
| **Surrogate** | *ENTITY_NAME* | Sara | Hannah | Maria |
| **Surrogate** | *ENTITY_NAME* | Sara | Lin | Maria |
| **PHI Similarity to Original** | Lowest | High | Low | Intermediate |
| **Maximum surrogate repeat size** | NA | NA | 1 | 3 |

**Surrogate Substitution Strategy Evaluation on UAB and MIMIC Corpora**

We evaluate the protective ability of surrogate addition by focusing on "*critical*" PHI as detailed in Supplementary Table 1 on the UAB Corpus, UAB Discharge Corpus and the MIMIC Corpus. We define "critical" such that a single FN (such as MRN) may alone be sufficient to identify a patient and necessitate immediate legal disclosure of the breach. A FN by de-identification software or humans in this critical category of PHI may be commonly identified even in a haphazard fashion by a casual reader of the document when the document employs no surrogate substitution (like MIMIC) or uses substituted text as shown in Table 1 under the Consistent strategy. However, a single FN would not be obvious to a reader if a Random or Markov strategy is used since the real name would not constitute the largest minority surrogate class. Thus, the ability to detect the real name as shown in Table 1 is dependent on the overall FNER, the distribution of critical entities and FN errors within those entities and the substitution strategy used. Using our UAB Corpus and the MIMIC corpus (described below), we simulate and assess 3 different resynthesis strategies for surrogate information: "Consistent, "Random" and "Markov" under 4 different FNER (0.1%, 0.5%, 1% and 5%). Leakage rates are computed at the document level such that a leak is considered to occur if the PHI signal (as measured by critical PHI FN count) exceeds the injected noise as measured by *maximum surrogate repeat size*. In practice this means that for the Consistent strategy where the same surrogate is repeated, any FN in a critical entity constitutes a document leak, whereas for the Random and Markov strategy this only occurs when FN errors exceed the *maximum surrogate repeat size*. A leak is defined at the patient level as one or more document leaks. Since results can vary depending on the distribution of FN errors in a corpus, we average our results over 1,000 FN error generation simulations using different random seeds. Our leakage calculations and simulations are available on Github[24] and were performed using Python version 3.9.

**Surrogate Substitution Strategy Evaluation on Synthetic Corpora**

We calculated the expected probability for leakage of critical information over a range of corpus sizes, ranging from 10 to 10,000 documents. For this calculation, we considered leakage to occur if the expected number of real patient entities (FN errors) in the corpus was greater than a set threshold for each of the substitution methods. For Consistent substitution, the threshold was set at 0, since even a single instance of a real patient entity can be readily distinguished from the single fake entity. For Random and Markov substitution, the threshold was set as the expected number of fake entities[25] based on a pool of 1,000 fake entities from which to randomly select using the transition probabilities described above. The threshold for Random substitution was equal to 1.015 and for Markov substitution was 2.028. We allowed the number of entities per document to assume the values of 15, 150, and 1500, and allowed the FNER to assume the values of 1% and 5%. We then calculated the expected number of real patient entities in the corpus as the product of the FNER times the number of entities per document times the number of documents. We used the binomial distribution to determine the p value for whether the expected number of real patient entities exceeded the threshold. Calculations were performed using R 4.1.2.

**UAB Corpus, UAB Discharge Corpus and MIMIC Corpus**

To assess the impact of surrogate substitution strategy on privacy protection, we evaluated BRATsynthetic on an in-house corpus of de-identified clinical text from 165 patients collected under IRB Protocol #300002331 (NIH Grant R01AG060993, Automating Delirium Identification and Risk Prediction in Electronic Health Records). Those notes were de-identified under IRB Protocol #300002212 "U-BRITE Deidentified Translational Data Repository for Research and Education at the University of Alabama" that we previously used and described for a gout flare corpus[26] to generate .ann files as input for BRATsynthetic. The corpus includes all available EHR notes for those patients, specifically 3,617 documents including 1,489,362 critical entities from both inpatient and outpatient encounters from 2014 to 2021. The UAB Discharge corpus is a subset of the main UAB Corpus containing only discharge summaries.
The MIMIC Corpus is a set of MIMIC discharge summaries that was obtained as part of participation in SemEval 2014 Task 7 "Analysis of Clinical Text" [27] and are sourced from outside UAB[28]. Supplementary Table 3 provides corpus statistics on the critical entity distributions of these corpora.

**BRATsynthetic Runtime Experiment**

Runtime performance was evaluated on a superset of the UAB corpus consisting of 28,547 documents that have been de-identified using a BERT based de-identification tool [29], but not been manually reviewed. Corpus statistics and experimental parameters are in Supplementary Table 4.

# RESULTS

## Surrogate Replacement Strategy Assessment
Figure 1. Impact of Surrogate Substitution Strategies on PHI Leakage on UAB Corpus

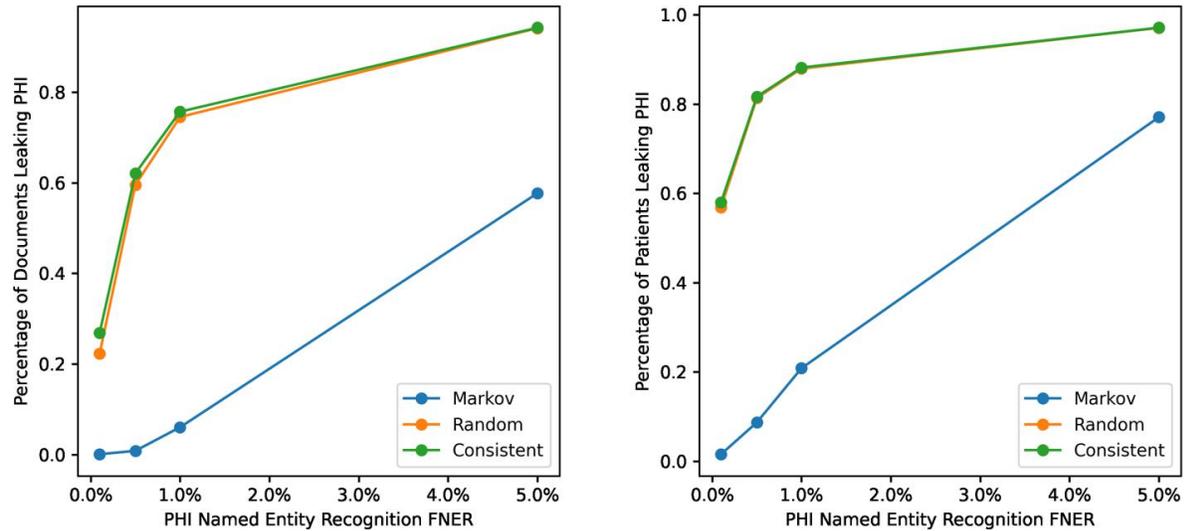

We compute PHI leakage rates as shown in Figure 1 for UAB Notes at the document (left) and patient (right) level. The Markov strategy shows a lower document leakage rate. At a lower FNER (1% or less) its impact is more pronounced, with leakage eliminated at the lowest error rate. Even at a 5% error rate the Markov strategy still performs substantially better than the two other strategies, which leak PHI close to 100% of the time.

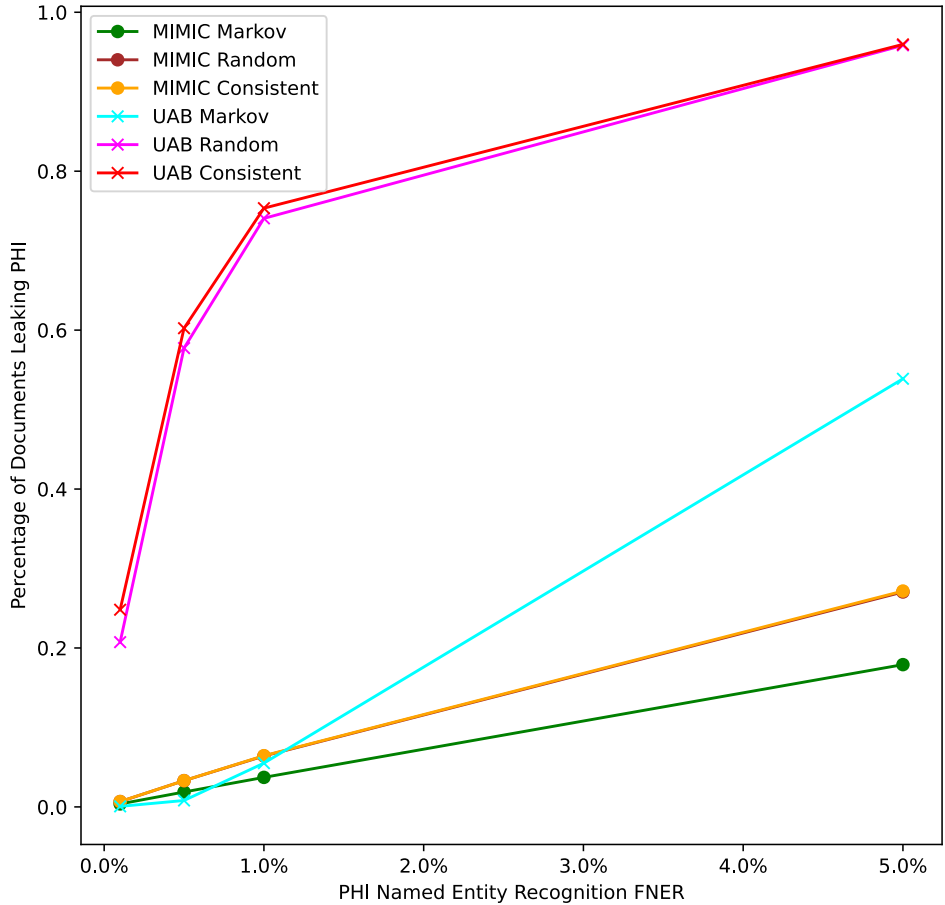

Figure 2. UAB vs MIMIC Corpus PHI Document Leakage

Figure 2 compares PHI leakage between the UAB and MIMIC corpus at the document level for discharge summaries only and indicates a similar substitution strategy performance between the 2 corpora, but dramatically lower PHI leakage rates for the MIMIC corpus which contains significantly fewer critical PHI entities per discharge summary relative to UAB.

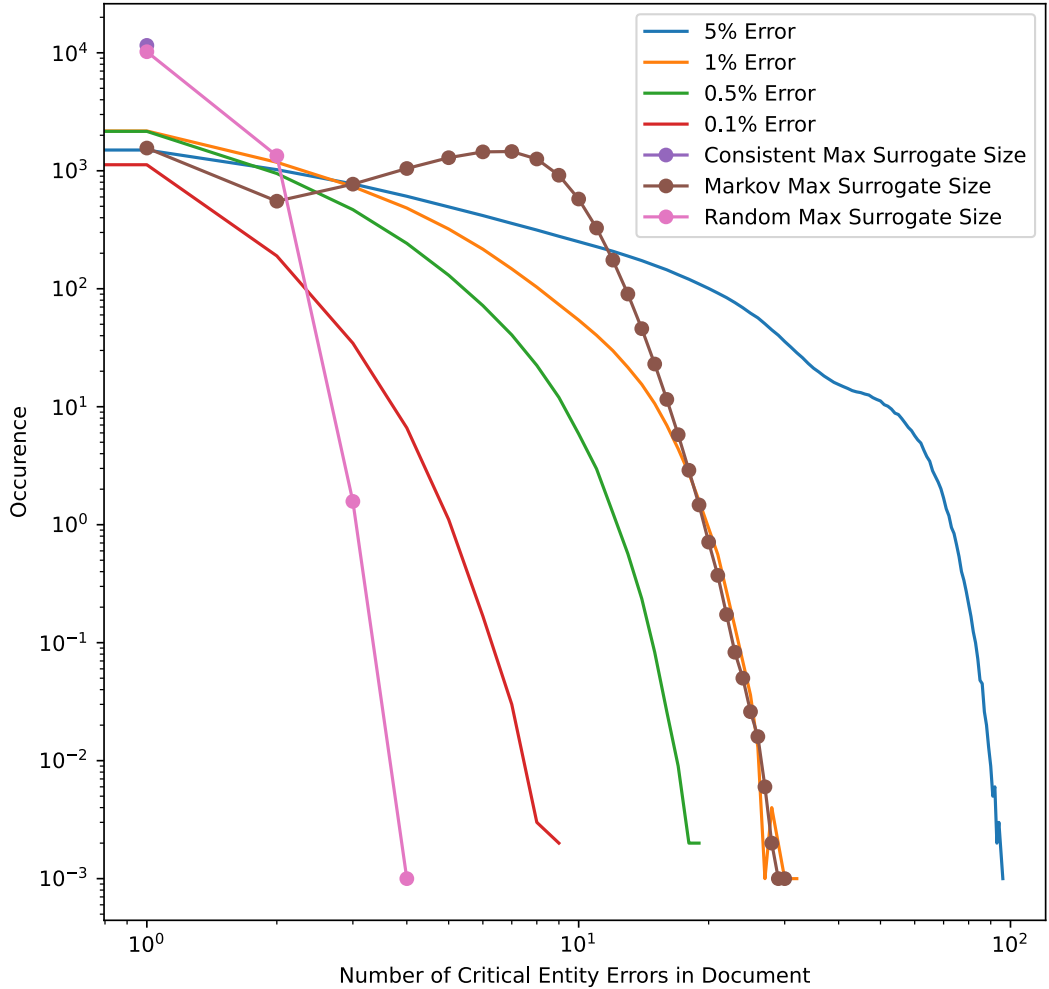

Figure 3. Markov Model Error Masking Simulation on UAB Corpus

Figure 3 shows the distribution of errors across 1,000 simulations, superimposed over the distribution of the *maximum surrogate repeat size* for all 3 substitution strategies. Consistent substitution always reutilizes the same surrogate replacement value for each class of critical PHI, so the *maximum surrogate repeat size* is always one. The Random substitution strategy allows for same surrogate replacement, but since the critical entity surrogate pool size is set at 1,000 almost all entities use a unique surrogate. Thus, for Random the *maximum surrogate repeat size* is usually 1, but it some cases it ranges from 2 to 4. The Markov strategy utilizes a state transition probability of 0.5 and so will generate a higher *maximum surrogate repeat size*. As shown in Figure 3, it will generally "cover" error rates of 1% or less in our data set.

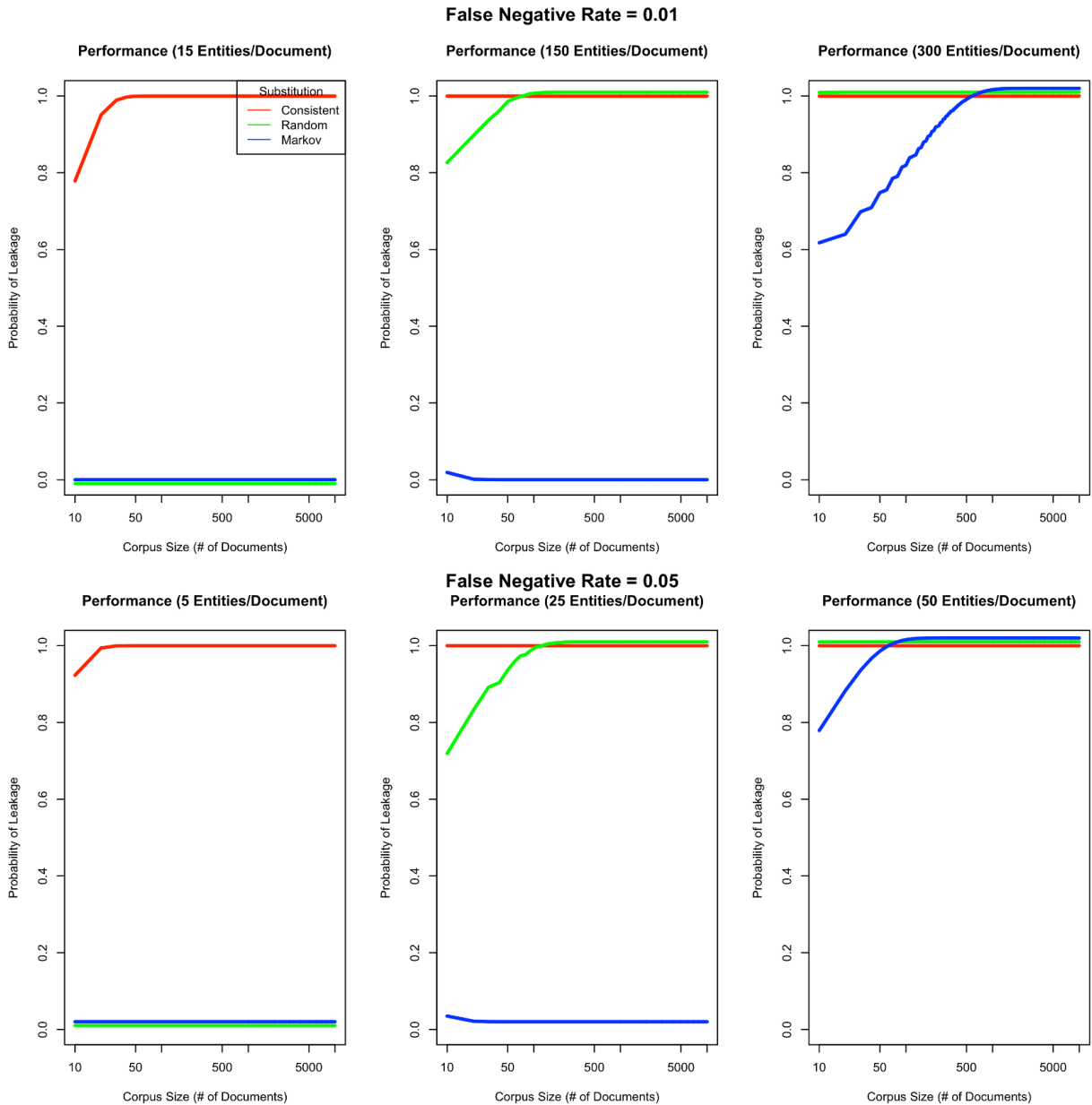

Figure 4. PHI Leakage on Synthetic Corpora

The top 3 charts of Figure 4 show the probability of PHI leaking given a 1% PHI FNER and varying input corpus sizes with either 15, 150 or 300 entities (critical or otherwise) per document. The bottom 3 charts of Figure 4 assume a higher FNER of 5% with 5, 25 or 50 entities per document. Although leakage results are reported at the document level, patient level calculations would be the almost same but would use a pooled entity count given that multiple documents from the same patient would be aggregated.

**Runtime Evaluation**

BRATsynthetic took an average of 20.9 seconds of system to process the entirety of the 28,547 files of the UAB corpus as shown in Supplementary Table 4.

# DISCUSSION

## Markov Replacement Strategy Evaluation

As shown in Figure 1, the Markov replacement strategy can significantly reduce the chance of PHI leakage relative to a corpus released without substitution or using a traditional Consistent substitution strategy. This effect is also seen in the discharge summaries of both the UAB Discharge and MIMIC corpora (Figure 2), with the effect much more pronounced on the MIMIC corpora which shows almost no leakage at lower error rates. This is due to the lower number of critical PHI entities contained in MIMIC corpus versus UAB Discharge because the format of current discharge notes at UAB includes a footer on every note page containing critical PHI including patient name and MRN. Tailored de-identification applications should be able to remove structured PHI with regular expressions yielding a PHI distribution similar to the MIMIC corpus.

The Markov substitution strategy is effective because it adds a variable amount of noise to a document data set (see Figure 3), at levels exceeding any original PHI signal that may not have been removed by the de-identification process. This shape of the Markov substitution strategy curve in Figure 3 is controlled by the state transition rate, which determines how often a surrogate is repeated. This could be adjusted in future work to reflect the anticipated number of FN errors in any entity to better conceal PHI leakage.

Figure 4 shows that for a range of PHI entity counts and FNERs the Markov substitution strategy allows for a larger released corpus size with the same or lower PHI leakage risk relative to Consistent or Random substitution strategies. While our results in Figures 1 to 3 focus on critical PHI, the protective effect shown in Figure 4 applies to any PHI entity (critical or not). Our focus on critical PHI versus non-critical PHI (such as the profession of a patient) reflects data availability in notes (replicated in other corpora) and the difficulty in assessing the impact or liability of inadvertent non-critical PHI release. Finally, while de-identification performance rivals human annotators[30], we do not yet suggest that a combination of PHI NER and Markov substitution strategy is sufficient to release a corpus without human oversight. In previous work [29] we have found that identifying descriptions of such as unusual accidents or rare occupations could leak PHI and are challenging to detect.

Our results in Figures 1, 2 and 4 indicate show that the implementation of an effective surrogate replacement strategy using Markov models can have a greater impact on the protection of patient privacy than marginal gains in PHI NER performance provided by additional training data. For example, Dernoncourt et al [31] show increased performance for artificial neural network and conditional random field de-identification algorithms with additional training data, but the total improvement as measured by F1 score show a maximum improvement difference of only about 2% between using 5% of available training data versus using all the available training data for the MIMIC data set. Our results indicate that surrogate replacement strategy can have a higher impact in protecting PHI versus than even a 4% improvement in recall (from 5% to a 1%) in PHI NER software, dependent on the total amount and distribution of PHI.

Finally, expanding strategies beyond Consistent substitutions could correct bias in notes. For example, the presumed gender of a name could be balanced in medical conditions that under-reported or under-documented in a particular gender so that machine learning algorithms do not learn a gender reporting bias. We leave bias correction to future work.

**BRATsynthetic**

BRATsynthetic uncouples surrogate generation from PHI NER and can thus leverage any independent improvements in PHI NER. The tradeoff is that PHI NER software is best positioned to recognize an appropriate substitution (such as one that maintains non-standard date formatting or even typos) whereas BRATsynthetic relies on PHI categories or its own regular expression-based interpretation of the text (if available) to generate the surrogate.

**Limitations**

Date shifting and synthetic clinical text can remove coreferential integrity as a side effect, our Markov and Random strategy makes such a removal explicit. We anticipate the impact would be minimal on tasks other than coreference resolution and leave its assessment on downstream tasks to future work.

Finally, both the faker library and BRATsynthetic have their source code available, which could potentially assist an attacker since some synthetic surrogates are generated from lists including patient and location names. This could be mitigated by corpus providers by the inclusion of held-out values in BRATsynthetic. Other forms of surrogate values such as dates and randomly generated identifiers consisting of alphanumeric text would not have this same vulnerability.

**CONCLUSION**

We show that a Markov substitution strategy can reduce PHI leakage on 2 real world corpora and on synthetic PHI distributions using a range of realistic false negative PHI NER rates and entity distributions and make our software freely available.

**ACKNOWLEDGEMENTS**

We would like to thank Ahana Chatterjee and Mojisola Fasokun for de-identification of the delirium corpus and the FAKER project which BRATsynthetic leverages for surrogate generation and Dr. Maria (Maio) Danila for assistance with editing the paper.

**DATA AVAILABILITY**
MIMIC and UAB PHI distributions used for all calculations will be released to Github[24] upon paper acceptance.

**FUNDING STATEMENT**
We acknowledge the support of 5P30AR072583 "Building and InnovatinG: Digital heAlth Technology and Analytics (BIGDATA)" and 5R01AG060993 "Automating Delirium Identification and Risk Prediction in Electronic Health Records" for the generation of the de-identified corpus.

**COMPETING INTEREST STATEMENT**
There are no competing interests.

**CONTRIBUTORSHIP STATEMENT**

JO wrote the first draft of the paper and all authors assisted with critical revisions and writing of the manuscript for important intellectual content. TO implemented BRATsynthetic with contributions from JO and AN. SA and RK performed statistical analysis.

**REFERENCES**

1. Li J, Zhou Y, Jiang X, et al. Are synthetic clinical notes useful for real natural language processing tasks: A case study on clinical entity recognition. Journal of the American Medical Informatics Association 2021;**28**(10):2193-201.
2. Lewis M, Liu Y, Goyal N, et al. Bart: Denoising sequence-to-sequence pre-training for natural language generation, translation, and comprehension. arXiv preprint arXiv:1910.13461 2019.
3. Raffel C, Shazeer N, Roberts A, et al. Exploring the limits of transfer learning with a unified text-to-text transformer. arXiv preprint arXiv:1910.10683 2019.
4. Brown TB, Mann B, Ryder N, et al. Language models are few-shot learners. arXiv preprint arXiv:2005.14165 2020.
5. Dernoncourt F, Lee JY, Uzuner O, Szolovits P. De-identification of patient notes with recurrent neural networks. J Am Med Inform Assoc 2017;**24**(3):596-606 doi: 10.1093/jamia/ocw156.
6. Neamatullah I, Douglass MM, Li-wei HL, et al. Automated de-identification of free-text medical records. BMC medical informatics and decision making 2008;**8**(1):1-17.
7. Schneble CO, Elger BS, Shaw DM. Google's Project Nightingale highlights the necessity of data science ethics review. EMBO Mol Med 2020;**12**(3):e12053 doi: 10.15252/emmm.202012053 [published Online First: 20200217].
8. El Emam K, Jonker E, Arbuckle L, Malin B. A systematic review of re-identification attacks on health data. PloS one 2011;**6**(12):e28071.
9. Heider PM, Obeid JS, Meystre SM. A comparative analysis of speed and accuracy for three off-the-shelf de-identification tools. AMIA Summits on Translational Science Proceedings 2020;**2020**:241.
10. Steinkamp JM, Pomeranz T, Adleberg J, Kahn Jr CE, Cook TS. Evaluation of automated public de-identification tools on a corpus of radiology reports. Radiology: Artificial Intelligence 2020;**2**(6):e190137.
11. Guzman B, Metzger I, Aphinyanaphongs Y, Grover H. Assessment of amazon comprehend medical: Medication information extraction. arXiv preprint arXiv:2002.00481 2020.
12. Clinacuity. Clindeid. Secondary Clindeid 2022. https://www.clinacuity.com/clinideid/.
13. An Easy-to-Use Clinical Text De-identification Tool for Clinical Scientists: NLM Scrubber. AMIA; 2015.
14. Aberdeen J, Bayer S, Yeniterzi R, et al. The MITRE Identification Scrubber Toolkit: design, training, and assessment. International journal of medical informatics 2010;**79**(12):849-59.
15. Computer-assisted de-identification of free text in the MIMIC II database. Computers in Cardiology, 2004; 2004. IEEE.
16. HIDE: an integrated system for health information DE-identification. 2008 21st IEEE International Symposium on Computer-Based Medical Systems; 2008. IEEE.
17. Johnson AE, Pollard TJ, Shen L, et al. MIMIC-III, a freely accessible critical care database. Scientific data 2016;**3**(1):1-9.
18. Carrell D, Malin B, Aberdeen J, et al. Hiding in plain sight: use of realistic surrogates to reduce exposure of protected health information in clinical text. J Am Med Inform Assoc 2013;**20**(2):342-8 doi: 10.1136/amiajnl-2012-001034 [published Online First: 20120706].
19. Carrell DS, Malin BA, Cronkite DJ, et al. Resilience of clinical text de-identified with "hiding in plain sight" to hostile reidentification attacks by human readers. J Am Med Inform Assoc 2020;**27**(9):1374-82 doi: 10.1093/jamia/ocaa095.
20. Carrell DS, Cronkite DJ, Li MR, et al. The machine giveth and the machine taketh away: a parrot attack on clinical text deidentified with hiding in plain sight. J Am Med Inform Assoc 2019;**26**(12):1536-44 doi: 10.1093/jamia/ocz114.
21. BRAT: a web-based tool for NLP-assisted text annotation. Proceedings of the Demonstrations at the 13th Conference of the European Chapter of the Association for Computational Linguistics; 2012.
22. Stubbs A, Uzuner Ö. Annotating longitudinal clinical narratives for de-identification: The 2014 i2b2/UTHealth corpus. J Biomed Inform 2015;**58 Suppl**:S20-S29 doi: 10.1016/j.jbi.2015.07.020 [published Online First: 20150828].
23. Faker Project. Secondary Faker Project. https://faker.readthedocs.io/en/master/.
24. BRATsynthetic [program]. 0.3 version. https://github.com/uabnlp/BRATsynthetic: Github, 2022.

25. Sigman K. Expected number of visits of a finite state Markov chain to a transient state. 2016; 2022(September 3, 2022). http://www.columbia.edu/~ks20/4106-18-Fall/Notes-Transient.pdf.
26. Osborne JD, Booth JS, O'Leary T, et al. Identification of Gout Flares in Chief Complaint Text Using Natural Language Processing. AMIA Annual Symposium Proceedings: American Medical Informatics Association, 2020:973.
27. SemEval-2014 Task 7: Analysis of clinical text. SemEval@ COLING; 2014.
28. Johnson A, Bulgarelli, L., Pollard, T., Horng, S., Celi, L. A., & Mark, R. MIMIC-IV. 0.4 ed. https://physionet.org, 2020.
29. Osborne JD, O'Leary, T., Mudano, A., Booth, J., Rosas, G., Peramsetty, G. S., Knighton, A., Foster, J., Saag, K., & Danila, M. I. . Gout Emergency Department Chief Complaint Corpora. 1.0 ed. Physionet, 2020.
30. Deleger L, Molnar K, Savova G, et al. Large-scale evaluation of automated clinical note de-identification and its impact on information extraction. Journal of the American Medical Informatics Association 2013;**20**(1):84-94.
31. Dernoncourt F, Lee JY, Uzuner O, Szolovits P. De-identification of patient notes with recurrent neural networks. Journal of the American Medical Informatics Association 2017;**24**(3):596-606.

**Supplementary Table 1. BRATsynthetic Safe Harbor Element Replacement**

| HIPAA Safe Harbor Category | Type | Description | Critical PHI | BRATsynthetic Replacement Type | MIMIC Identified Type |
|---|---|---|---|---|---|
| NAMES | DOCTOR | Health care provider name | No | DOCTOR | DOCTOR |
|  | PATIENT | Patient name | **Yes** | PATIENT | PATIENT |
|  | USERNAME | User IDS of providers | No | USERNAME | - |
| GEO LOCATION | LOC-OTHER | Identifiable locations and landmarks | No | LOCATION-OTHER | LOC-OTHER |
|  | HOSPITAL | Hospital or clinic name | No | HOSPITAL | HOSPITAL |
|  | WARD | Ward or unit name | No | - | WARD |
|  | ZIP | Zip code | No | ZIP | ZIP |
|  | ORGANIZATION | Employers | No | ORGANIZATION | ORGANIZATION |
|  | COUNTRY | Country | No | COUNTRY | COUNTRY |
|  | STATE | State or province name | No | STATE | STATE |
|  | CITY | Name of city | No | CITY | LOC-OTHER |
|  | STREET | Street address | No | STREET | STREET |
| DATES | DATE | Year | No | DATE (regex) | DATE |
|  |  | Month/Day |  | DATE (regex) | DATE |
|  |  | Day of the week | No | DATE (regex) | - |
|  | HOLIDAY | Holidays | No | DATE (regex) | HOLIDAY |
|  | AGE | AGE >= 90 | No | AGE | AGE_90_ANDUP |
|  |  | AGE < 90 | No | AGE | - |
| PHONE | PHONE | Telephone numbers | **Yes** | PHONE | PHONE |
| VEHICLE IDS | VEHICLE_ID | Vehicle identification number or license | **Yes** | IDNUM (regex) | IDNUM |
| FAX | FAX | Fax numbers | **Yes** | PHONE | PHONE |
| DEVICE IDS | DEVICE IDS | Device identifiers | **Yes** | DEVICE (regex alphanumeric) | IDNUM |

| | | and serial numbers | | | |
|---|---|---|---|---|---|
| IDNUM | IDNUM | License and health plan numbers | **Yes** | IDNUM (regex) | |
| MEDICAL RECORD | MEDICAL RECORD | Medical record number | **Yes** | IDNUM (alphanumeric) | |
| SSN | SSN | Social security numbers | **Yes** | IDNUM (regex) | SSN |
| ACCOUNT ID | ACCOUNT ID | Account numbers | **Yes** | IDNUM (alphanumeric) | ACCOUNT ID |
| EMAIL | EMAIL | Email address | **Yes** | EMAIL | - |
| URL | URL | URL | No | URL | - |
| BIOMETRIC ID | BIOID | Biometric identifiers, including finger and voice prints | NA | BIOID (alphanumeric) | |
| IP ADDRESS | IP ADDRESS | Internet Protocol Address | No | URL | - |

**Supplementary Table 2. Differences Between Safe Harbor Elements and BRATsynthetic**

| Safe Harbor Element | BRATsynthetic Category |
|---|---|
| IMAGE | Images are not de-identified |
| UNIQUE | Unique identifying phrases must be manually redacted |
| Not covered under safe harbor | PROFESSION (also a category in I2B2) |
| Not covered under safe harbor | TIME (exclusive to BRATsynthetic) |

**Supplementary Table 3. UAB and MIMIC Critical Entity Distribution**

| | Document Statistics | | | Patient Statistics | |
|---|---|---|---|---|---|
| | UAB Total | UAB Discharge | MIMIC Discharge | | UAB Total |
| Critical Entities Mean | 388.5 | 355.6 | 6.8 | Critical Entities Mean | 8123.0 |
| Critical Entities Median | 224 | 199 | 5 | Critical Entities Median | 985 |
| Critical Entities Range | 2 – 2545 | 10-2414 | 2-76 | Critical Entities Range | 7 - 321945 |

**Supplementary Table 4. BRATsynthetic Runtime Experiment**

| Machine | Documents | Words | PHI Entities | Runtime (5 run average) |
|---|---|---|---|---|
| 3.4 Ghz Quad-Core Intel Core i7 CPU with 32 GB 1600 MHz DDR3 | 28,547 | 32,432,577 | 1,710,386 | 20.9 seconds system time |